\documentclass[aps,prl,12pt]{revtex4-1}

\usepackage{amssymb,amsmath}
\usepackage{graphicx}

\begin{document}

\title{Determination of the Parameters of a \\
Color Neutral 3D Color Glass Condensate Model}

\author{\c{S}ener \"{O}z\"{o}nder}

\email{ozonder@umn.edu}

\affiliation{School of Physics and Astronomy, University of Minnesota, Minneapolis,
Minnesota 55455, USA}

\begin{abstract}
We consider a version of the McLerran-Venugopalan model 
by Lam and Mahlon
where confinement is implemented via colored noise in
the infrared. 
This model does not assume an infinite momentum frame, hence the boosted nuclei are not infinitely thin. Instead, the nuclei have a finite extension in the longitudinal direction and therefore depend on the longitudinal coordinate.
In this fully three dimensional framework an x dependence of the 
gluon distribution function emerges naturally. 
In order to fix the parameters of the model, we calculate the gluon
distribution function and compare it with the JR09 parametrization
of the data. 
We explore the parameter space of the model to attain
a working framework that can be used to calculate the initial conditions
in heavy ion collisions.
\end{abstract}
\maketitle


\section{Introduction}

In heavy ion collisions at very high energies, 
soft (small-x) partons, mostly
gluons, constitute a big portion of 
the wave function of the colliding
nuclei. Since the occupation number of 
the gluons is very high, they
can be treated in the framework of 
classical Yang-Mills theory. When
two nuclei pass through each other, 
the classical nonabelian fields from each
nucleus interact with each other and form color flux tubes between
the receding nuclei, which ultimately lead to the quark gluon plasma. 
The energy density distribution of the initial state 
determines the multiplicity and spectrum
of the final particles seen in the detectors. 
The initial energy density distribution
is an input to the hydrodynamics description 
of the evolving quark gluon
plasma and, in principle, it can be calculated 
ab initio from Quantum Chromodynamics
(QCD).

The energy initially deposited in the color flux tubes can be related to 
the two-point vector potential correlation function
 $\langle A_{i}^{a}(\boldsymbol{x})A_{i}^{b}(\boldsymbol{x}^{\,\prime})\rangle$ where 
the setup may be two (transverse) or three dimensional, 
depending on whether or not the Lorentz contracted nucleus is assumed to have a longitudinal
thickness in the lab frame. This correlator can be derived analytically from the color
charge density correlator 
$\langle\rho^{a}(\boldsymbol{x})\rho^{b}(\boldsymbol{x}^{\,\prime})\rangle$
by using the classical Yang-Mills equations.

The color charge density of a nucleus $\rho^{a}(\boldsymbol{x})$
during a collision cannot be known; 
on the other hand, the fluctuations in the
color charge density can be studied in the effective field theory approach
with ensemble averaging.
In the Color Glass Condensate (CGC), a framework for  
slowly evolving high density gluons within the ultrarelativistic nucleus,
the fast partons are seen as sources
of the small-x (soft) gluon radiation, where x 
is the
fraction of the total longitudinal momentum carried by a parton.
After integrating out the fast partons, 
the observables can be calculated by averaging them over the ensemble of all 
possible color charge configurations. Once the correlator
$\langle\rho^{a}(\boldsymbol{x})\rho^{b}(\boldsymbol{x}^{\,\prime})\rangle$ 
is specified, 
it can be linked to the vector field correlator 
$\langle A_{i}^{a}(\boldsymbol{x})A_{i}^{b}(\boldsymbol{x}^{\,\prime})\rangle$
from which the gluon distribution function $xg(x,Q^{2})$ and other observables,
such as the initial energy density,
can be calculated. Here $\langle \cdots \rangle$ 
denotes the ensemble average. See Ref. \cite{Iancu:2003xm} 
for a comprehensive review.

In the early formulation of the CGC by McLerran and Venugopalan
\cite{McLerran:1993ni,McLerran:1993ka,
McLerran:1994vd,Ayala:1995kg,Ayala:1995hx},
the spectrum of the Gaussian color fluctuations was taken to be
white noise. This results in arbitrarily long wavelength fluctuations where 
$\langle A_{i}^{a}(\boldsymbol{x})A_{i}^{b}(\boldsymbol{x}^{\,\prime})\rangle$
diverges in the infrared. This problem originates from the fact that 
confinement effects (color neutrality) have not been taken into
account. Adding a gluon mass to the gluon
propagator so as to bypass this problem breaks gauge invariance 
which is required later to convert 
the solution of the classical Yang-Mills equations
from the axial gauge to the light-cone gauge by means of Wilson lines.

In the original McLerran-Venugopalan (MV) model, the colliding nuclei
are considered
to be two-dimensional infinitely thin sheets traveling at the speed of light. 
In this approximation,
the model does not depend on the longitidunal coordinate, 
hence there is no dependence on the momentum fraction x
of a given parton.
In other words, all of the partons in the nucleus have
the same momentum fraction x. 
Being an artifact of the infinite momentum
frame, the lack of x dependence in the model does not reflect the
true nature of the x dependent gluon distribution functions. 

In this paper, we consider a color neutral three dimensional 
McLerran-Venugopalan (3dMVn) model 
by Lam and Mahlon \cite{Lam:1999wu,Lam:2000nz},
where the spectrum of the Gaussian fluctuations is taken to be infrared-safe
colored noise. Colored noise creates a correlation hole 
in the correlation function on the size
of a nucleon and this leads to color neutrality.
(The term ``colored noise'' is not related to the color
charge of QCD.)
 This model comes with another important 
 feature that the incoming nuclei are
not exactly on the light cone. Longitudinal coordinate dependence
produces an x dependent gluon distribution function. This enables us to
compare the model with the data parametrization of gluon distribution
functions and therefore fix 
the parameters of the 3dMVn model. Once this is achieved, 
the 3dMVn model can be used to calculate the initial energy 
density distribution in heavy ion collisions.

The MV model provides a framework for calculating 
the vector field correlation function 
$\langle A_{i}^{a}(\boldsymbol{x})A_{i}^{b}(\boldsymbol{x}^{\,\prime})\rangle$
from which the gluon distribution function as well as the 
initial energy density distribution 
can be calculated.
We emphasize that our ultimate goal is calculating the latter, 
which will be pursued in a follow-up paper.
In this paper we focus on the calculation of gluon distribution function
and its comparison with data for the purpose of fixing 
the free parameters
in the 3dMVn model.

The outline of the paper is as follows.
First, we give a brief overview of the color neutral and x-dependent version of the McLerran-Venugopalan model (3dMVn) given by
Lam and Mahlon \cite{Lam:1999wu,Lam:2000nz}. Next, we calculate
the gluon distribution function from the 3dMVn model.
This model works at low-$Q^{2}$
where no data for the gluon distribution functions is available. 
For that reason, we evolve our results
with the Dokshitzer-Gribov-Lipatov-Altarelli-Parisi (DGLAP) equation to higher momenta where data is available. 
This is followed by a discussion of the data
parametrization that we use to compare with the 3dMVn model.
Lastly, we compare the 3dMVn
with the data parametrization.
The final section summarizes our analysis and includes a
discussion regarding how the results will be used in 
the future to calculate initial
energy density distribution.


\section{Color Neutrality}

In the original two dimensional MV model, the average and the fluctuations
of the color charge density of a nucleus are given by 
\begin{eqnarray}
\langle\rho^{a}(\boldsymbol{x}_\perp^{~} )\rangle & = & 0,\label{eq:vev}\\
\langle\rho^{a}(\boldsymbol{x}_\perp^{~} )
\rho^{b}(\boldsymbol{x}^{\,\prime}_{\perp})\rangle & = & 
\delta^{ab}\mu_{A}^{2} \, {\cal D}({\boldsymbol{x}_\perp^{~} }
-\boldsymbol{x}^{\,\prime}_{\perp}),\label{eq:variance}
\end{eqnarray}
where $\boldsymbol{x}_\perp^{~} $ is the 
transverse coordinate system for the 
infinitely thin nucleus, $\mu_{A}^{2}$ is the 
average color charge density squared 
per unit (transverse) area for a nucleus, and
$\langle\cdots\rangle$ denotes the ensemble average. 
The function
${\cal D}$ determines the spectrum of the fluctuations. 
When white noise
is assumed, ${ {\cal D}(\boldsymbol{x}_\perp^{~} 
-\boldsymbol{x}^{\,\prime}_{\perp})=\delta^{2}(\boldsymbol{x}_\perp^{~} 
-\boldsymbol{x}^{\,\prime}_{\perp}) }$,
no correlation occurs between different points in a nucleus. This
also means that fluctuations in each momentum mode, including the
zero mode, are equally likely and there is no characteristic scale. 
In this case, even though the correlator 
$\langle A_{i}^{a}(\boldsymbol{x}_\perp^{~})A_{i}^{b}
(\boldsymbol{x}^{\,\prime}_{\perp})\rangle$
is calculated with an infrared cutoff  $\Lambda_{\rm{QCD}}$ 
\cite{JalilianMarian:1996xn}, it still diverges like 
$(\boldsymbol{x}_\perp^2)^{\boldsymbol{x}_\perp^2}$ in the infrared. 
Therefore, the Fourier transform
of it, which
is necessary to calculate the gluon distribution function, 
does not exist \cite{Lam:1999wu}.
This cut-off can also be understood as a gluon mass in 
the gluon propagator
in the form 
$  ({\boldsymbol{q}_\perp^{2}+m_{\rm{gluon}}^{2}})^{-1} $.
It explicitly breaks
the gauge invariance that is needed to convert the solutions of
classical Yang-Mills equations from the axial gauge to 
the light-cone gauge
where Wilson lines are to be used.

In reality, a nucleus is color neutral 
on scales larger than a
nucleon size. A color neutral correlation function
can be derived from a simple model of a nucleus
where nucleons are composed of quark and antiquark 
pairs \cite{Kovchegov:1996ty}.
If we consider a two dimensional nucleus for a moment,
the assumption that only the quark and antiquark pair from the same
nucleon can interact with each other produces a correlator of the
form \cite{Lam:1999wu}
\begin{equation}
\langle\rho^{a}(\boldsymbol{x}_\perp^{~} )\rho^{b}
(\boldsymbol{x}^{\,\prime}_{\perp})\rangle=\delta^{ab}\mu_{A}^{2}
\left[\delta^{2}(\boldsymbol{x}_\perp^{~} -\boldsymbol{x}^{\,\prime}_{\perp})-
\frac{\exp\left[-\vert \boldsymbol{x}_{\perp}^{~}
-\boldsymbol{x}^{\,\prime}_{\perp}\vert^{2}
/ 4 \lambda^{2} \right]}{4\pi \lambda^{2}}
\right].\label{eq:colorednoise}
\end{equation}
Here the parameter $\lambda$, which
we will refer to as a correlation length, 
is of the order of 
a nucleon size $\sim1\,\rm{fm}$.
This parameter takes away the need for a sharp 
infrared cutoff as it is used in the original MV treatment; moreover,
it makes the Fourier transform of the vector field correlation function
well-defined and it renders the model infrared safe.
In Eq. (\ref{eq:colorednoise}), decorrelation 
due to the white noise ${\cal D}(\boldsymbol{x}_\perp^{~} 
-\boldsymbol{x}^{\,\prime}_{\perp})=\delta^{2}(
\boldsymbol{x}_\perp^{~} -\boldsymbol{x}^{\,\prime}_{\perp})$
is modified by the last term for distances 
$\vert \boldsymbol{x}_\perp^{~} -\boldsymbol{x}^{\,\prime}_{\perp}\vert
\lesssim \lambda$,
reflecting the assumption that there is a correlation between
the partons confined to a region smaller than the nucleon size. 
The second part in Eq. (\ref{eq:colorednoise}) removes the zero mode
$\vert \boldsymbol{q}_{\perp}^{~} \vert=0$ from 
the white noise spectrum, hence colored noise.
The Fourier transform of the
correlator in Eq. (\ref{eq:colorednoise}) is
proportional to 
$1-e^{-\lambda^{2}\boldsymbol{q}_{\perp}^{2}}$, 
which vanishes as 
$\vert \boldsymbol{q}_{\perp}^{~} \vert \rightarrow0$.
This makes the model infrared safe.

In this paper, we adopt a three dimensional correlator 
given by Lam and Mahlon \cite{Lam:2000nz}
\begin{equation}
\langle\rho^{a}(0,0)\rho^{b}(\boldsymbol{x})\rangle=\delta^{ab}
\kappa_{A}^{3}  
\left[\delta^{3}(\boldsymbol{x})-\frac{3}{4\pi \lambda^{2}}
\frac{
\exp \left( - \sqrt{3} \vert \boldsymbol{x} \vert / \lambda \right)
}{
\vert \boldsymbol{x} \vert
}
\right].\label{eq:Yukawa}
\end{equation}
where $\kappa_{A}^{3}$ is the three dimensional 
average color charge density
\begin{equation}
\kappa_{A}^{3}=\frac{3 A C_F}{N_{c}^{2}-1} 
\frac{1}{V}=\frac{3 A}{2 N_c} \frac{1}{V},
\label{kappa}
\end{equation}
Here $N_c$ is the number of colors, $V$ is the 
volume of a nucleus and the color factor
is defined as $C_F=(N_{c}^{2}-1)/(2 N_c)$.
We define $d^{3}\boldsymbol{x}\equiv dx_{\parallel}d^{2} 
\boldsymbol{x}_\perp^{~} $
where $x_{\parallel}$ is the longitudinal coordinate in the direction
of the beam axis, and $\boldsymbol{x}_\perp^{~}$ 
is the coordinate on the transverse
plane. The Fourier transform of the corralator 
(\ref{eq:Yukawa})
is given by
\begin{equation}
\widetilde{\langle\rho^{a}\rho^{b}\rangle}=\delta^{ab}
\kappa_{A}^{3}
\left[1-\frac{1}{1+ \lambda^{2} \boldsymbol{q}^{2}/3} \right].
\label{eq:Yukawa-Fouriered}
\end{equation}
Despite the slight difference between the correlators 
in Eqs. (\ref{eq:colorednoise})
and (\ref{eq:Yukawa}), they produce
similar results. It can be readily seen from 
Eq. (\ref{eq:Yukawa-Fouriered})
that the zero mode does not exist in the spectrum 
since $\widetilde{\langle\rho^{a}\rho^{b}\rangle}\rightarrow0$
as $\vert \boldsymbol{q} \vert \rightarrow0$.


\section{Longitudinal Dependence}

In ultrarelativistic heavy ion collisions, the infinite momentum frame
provides a good starting point for the parton picture of the nucleus.
However, the nucleus becomes infinitely thin when it is exactly
on the light cone and the gluon distribution function 
turns out to be independent of x. Here we review 
the formulation by 
Lam and Mahlon \cite{Lam:2000nz} for the
case where the nucleus is boosted to speed $\beta$.
In the lab frame, the thickness of a nucleus of radius $R$ 
becomes of the order of $R/\gamma$.

The current for
a nucleus moving in the $+z$ 
direction
is
\begin{equation}
J_{r}^{0}=\rho(-z,{\boldsymbol{x}_\perp^{~} }_{r});
\,\,\,\,\boldsymbol{J}_{r}=0, \label{eq:rest-current1}
\end{equation}
where the subscript ``r'' stands for the rest frame
and the negative sign in front of the
$z$ is for later convenience. With the redefinition 
$x^{\pm}=-x_{\mp}=(t\pm z)/\sqrt{2}$,
the current can be rewritten in the light-cone coordinates (still
in the rest frame),
\begin{equation}
J_{r}^{+}=J_{r}^{-}=\frac{1}{\sqrt{2}}\rho\left(
\frac{1}{\sqrt{2}}(x_{r}^{-}-x_{r}^{+}),
{\boldsymbol{x}_\perp^{~} }_{r}\right);\,\,\,\, 
{\boldsymbol{J}_\perp}_{r}=0.\label{eq:rest-light-cone}
\end{equation}

When we go from the rest frame to the lab frame where the nucleus
moves with speed $\beta$, the current in Eq. (\ref{eq:rest-light-cone})
becomes 
\begin{equation}
J^{+}=\frac{1}{\varepsilon}\rho\left(\frac{1}{\varepsilon}
x^{-}-\frac{\varepsilon}{2}x^{+},
\boldsymbol{x}_\perp^{~}  \right);\,\,\,\, 
J^{-}=\frac{\varepsilon}{2}J^{+};\,\,\,\,
{\boldsymbol{J}_\perp}=0,\label{eq:lab-light-cone}
\end{equation}
where 
\begin{equation}
\varepsilon=\sqrt{\frac{2(1-\beta)}{1+\beta}}.
\end{equation}
Here we can define a new longitudinal coordinate
\begin{equation}
 x_{\parallel}\equiv\frac{1}{\varepsilon}x^{-}
 -\frac{\varepsilon}{2}x^{+}, 
\label{newcoord}
\end{equation}
which is essentially the Lorentz
tranformation of $-z=\left( x^{-}-x^{+} \right) / \sqrt{2}$
(see the definition below Eq. (\ref{eq:rest-current1})). 
The Eq. (\ref{eq:lab-light-cone})
can be contrasted with the commonly used current 
in the infinite momentum
frame (imf) where the nucleus is infinitely thin,
\begin{equation}
J_{\rm{imf}}^{+}=\delta(x^{-})\rho(\boldsymbol{x}_\perp^{~} );\,\,\,\, 
J_{\rm{imf}}^{-}=0;\,\,\,\,
{\boldsymbol{J}_\perp}_{\rm{imf}}=0.
\end{equation}


\section{Calculation and Evolution of The Gluon Distribution Function}

We now turn to the calculation of the gluon 
distribution function from the 3dMVn model. 
Later we will compare it with data  
in order to fix the parameters $\alpha_s$ and $\lambda$.
The gluon distribution function of a nucleus with a 
baryon number $A$ 
can be expressed as an integral of the gluon number 
density over the transverse
momenta,
\begin{equation}
g_{A}(x,Q^{2})\equiv\int_{\vert  {\boldsymbol{q}_\perp} \vert
\leqslant Q}d^{2}{\boldsymbol{q}_\perp} \frac{dN}{dxd^{2}
\boldsymbol{q}_\perp}.\label{eq:gluon-distro}
\end{equation}
The gluon number density in the lab frame for a nucleus moving with speed
$ $$\beta$ is given as a Fourier transform of the two-point vector
field correlation function \cite{Lam:2000nz}
\begin{equation}
\frac{dN}{dq_{\parallel}d^{2} {\boldsymbol{q}_\perp}  }\equiv
\frac{q_{\parallel}}{4\pi^{3}}\int d^{3}\boldsymbol{x}
\int d^{3}\boldsymbol{x}^{\,\prime} 
e^{i \boldsymbol{q} \cdot(\boldsymbol{x}-\boldsymbol{x}^{\,\prime})}
\langle A_{i}^{a}(\boldsymbol{x})A_{i}^{a}(\boldsymbol{x}^{\,\prime})
\rangle,\label{eq:gluon-number-density}
\end{equation}
Here $q_{\parallel}$ is the momentum conjugate to the coordinate
defined in Eq. (\ref{newcoord}).
The Eq. (\ref{eq:gluon-number-density}) can be related to 
 $ dN/dxd^{2} \boldsymbol{q}_\perp$
by using the relation $x\equiv q_{\parallel} / m$,
which itself can be derived from the definition 
$x\equiv q^{+}/Q^{+}= \varepsilon q^{+}/m$.
Here $m$ is the nucleon mass, $q^{+}$ and $Q^{+}$ are the momenta
of the gluon and the nucleon.

Solving the classical Yang-Mills 
equations for the source in Eq. (\ref{eq:rest-light-cone})
in the covariant gauge and transforming the solution into the light-cone
gauge by using Wilson lines, one can express the 
correlator $ $$\langle A_{i}^{a}(\boldsymbol{x})A_{i}^{a}
(\boldsymbol{x}^{\,\prime})\rangle$
in Eq. (\ref{eq:gluon-number-density}) in terms of 
the color charge density
correlator $ $$\langle\rho^{a}(\boldsymbol{x})\rho^{a}
(\boldsymbol{x}^{\,\prime})\rangle$.
We skip the details of this calculation here and refer the reader
to the original paper \cite{Lam:2000nz}. Using the 
correlator in Eq. (\ref{eq:Yukawa}),
the gluon number density can be written as
\begin{equation}
\frac{dN}{dxd^{2} {\boldsymbol{q}_\perp}}=\frac{
8 A
\alpha_{s}}{\pi^{2}}\frac{1}{x}\int d^{2} {\boldsymbol{\Delta}_\perp}
e^{i {\boldsymbol{q}_\perp}\cdot
{\boldsymbol{\Delta}_\perp}}{\cal L}(x;{\boldsymbol{\Delta}_\perp})
{\cal E}(v^{2}L({\boldsymbol{\Delta}_\perp})),\label{eq:gluon-distro-LM}
\end{equation}
Here the integration is over the transverse coordinate ${\boldsymbol{\Delta}_\perp}=
\boldsymbol{x}_\perp^{~}  -\boldsymbol{x}^{\,\prime}_{\perp}$.
The functions $\cal{L}$ and $L$ are convolutions 
of the two gluon propagators
at two different points in the same nucleus 
connected by the three dimensional noise term
${\cal D}( \boldsymbol{x} - \boldsymbol{x}^{\,\prime})$. 
These two functions
can be seen as a pair distribution function. 
The nuclear correction factor
$\cal{E}$ takes into account the nuclear 
geometry and $v^2$ controls the strength
of dependence on this geometry.

The functions used in Eq. (\ref{eq:gluon-distro-LM}) 
are given as \cite{Lam:2000nz}
\begin{equation}
{\cal L} (x;{\boldsymbol{\Delta}_\perp})=-\frac{1}{12 \pi} 
(x m \lambda)^2 K_0(x m \Delta_\perp),
\end{equation}
and
\begin{equation}
L({\boldsymbol{\Delta}_\perp})=-\frac{\lambda^2}{6 \pi} 
\left[ K_0\left( 
\frac{\sqrt{3} \Delta_\perp}{\lambda}\right)+
\rm{ln}\left(\frac{\sqrt{3} 
\Delta_\perp}{2\lambda}\right)+\gamma_E    \right],
\end{equation}
where $\Delta_\perp=\vert \boldsymbol{\Delta}_\perp \vert$. In the calculations of Lam and Mahlon \cite{Lam:2000nz}, 
the nuclear matter density is taken to be uniform. 
The nuclear correction factor $\cal{E}$ is determined by
geometry of the nucleus. For cylindrical 
and spherical nuclei it is 
given by
\begin{equation}
{\cal E}(z)=
\begin{cases}
\displaystyle{\frac{1}{z}} (e^{z}-1) & \text{(cylindrical),} \\ \\
\displaystyle{\frac{3}{z^{3}}} [1-\frac{1}{2}z^{2}+e^{z}(z-1)] & 
\text{(spherical).}
\end{cases}
\end{equation}
This function appears in Eq. (\ref{eq:gluon-distro-LM}) as 
${\cal E}(v^2 L({\Delta_\perp}))$. 
The $v^2$ controls how much the results depend 
on the nuclear geometry; it is given by
\begin{equation}
v^2=
\begin{cases}
\displaystyle{ \frac{3 A g^4}{2 \pi R^2} } \approx 24 
\pi \alpha_s^2 A^{1/3} R_0^{-2}  & \text{(cylindrical),} 
\\ \\
\displaystyle{ \frac{9 A g^4}{4 \pi R^2} } \approx 36 
\pi \alpha_s^2 A^{1/3} R_0^{-2} & \text{(spherical).}
\end{cases}
\end{equation}
Here $R=R_0 A^{1/3}$ and we take $R_0=1\,\rm{fm}$.
The difference between the spherical and cylindrical nuclei
is negligibly small. In our calculations, we use the formulae for a cylindrical
nucleus. 

Now we are ready to calculate the gluon distribution function 
$xg_{A}(x,Q^{2})$
for a nucleus by substituting Eq. (\ref{eq:gluon-distro-LM}) in Eq.
(\ref{eq:gluon-distro}),
\begin{equation}
xg_{A}(x,Q^{2}) = \frac{8 A \alpha_{s}}{\pi^{2}}\int_{\vert
{\boldsymbol{q}_\perp}\vert\leqslant Q}d^{2}
{\boldsymbol{q}_\perp}\int d^{2}
{\boldsymbol{\Delta}_\perp} e^{i {\boldsymbol{q}_\perp}
\cdot {\boldsymbol{\Delta}_\perp}}
{\cal L}(x;{\boldsymbol{\Delta}_\perp}){\cal E}(v^{2}
L({\boldsymbol{\Delta}_\perp})).\label{eq:gluon-distro-final}
\end{equation}
The momentum integration in the equation above is
trivial.
Owing to the longitudinal coordinate $x_{\parallel}$, 
the fractional momentum distribution
function $xg_{A}(x,Q^{2})$ in Eq. (\ref{eq:gluon-distro-final}) 
is x dependent, in contrast
to the results of the original two dimensional MV model. 

The region of validity of the 3dMVn model within the 
two dimensional parameter space
$(x,Q^{2})$, likewise the original MV model, is restricted 
by the assumptions made regarding
the weak coupling limit, color coherence and 
color averaging \cite{Lam:2000nz}. 

The lower limit of x is set for a gold nucleus by
\begin{equation}
x \sim \frac{A^{-1/3}}{m R_0} \simeq 0.035
\label{lowerlim}
\end{equation}
where $R_0$ is the nucleon size 
$\sim 1\,\rm{fm}$. This is the longitudinal momentum 
fraction at which the gluons start to resolve the Lorentz 
contracted thickness of the	
nucleus.

The weak coupling limit imposes an upper
limit on x such that
\begin{equation}
x\lesssim\frac{4}{\pi m R_0}\simeq0.25.
\end{equation}
For very large $Q^2$,
not enough color charge would be seen, hence the 
Gaussian approximation for
the color charge fluctuations $\langle\rho^{a}(\boldsymbol{x})
\rho^{b}(\boldsymbol{x}^{\,\prime})\rangle$
would not be valid. Similarly, at the scale of color neutral nucleons
where $Q^{2}$ is very small, again there would not be
enough color charge to average over. 
These considerations put limits
on the momentum scale 
\begin{equation}
\frac{\pi m }{4 R_0}x\lesssim
Q^{2}\lesssim\frac{4 }{m R_0^3}\frac{1}{x}.
\label{Qvalidity}
\end{equation}
\begin{figure}[t]
\begin{centering}
\includegraphics[scale=0.84]{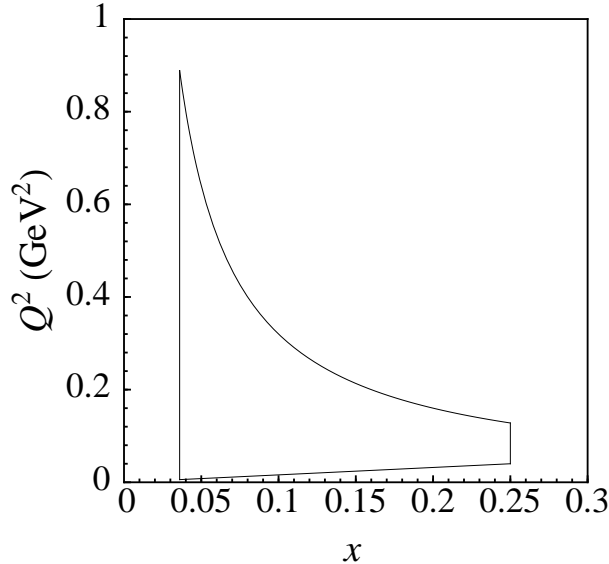}
\par\end{centering}
\caption{Approximate region of validity of 
the 3dMVn model for gold nucleus (A=197). }
\label{figure:validity}
\end{figure}
For $R_0=1\,\rm{fm}$, Eq. (\ref{Qvalidity}) becomes
\begin{equation}
0.16\, x\lesssim Q^{2}\lesssim\frac{0.032}{x}.
\end{equation}

In our calculations, we will take { $Q^{2}=0.55\,{\rm GeV^{2}}$}.
The reason for this choice is that it is at
the upper limit of validity of the 3dMVn model 
and at the lower
limit of the parton distribution function to which we will compare.
While the model spans a 
larger range in x for $Q^2 < 0.55\,\rm{GeV^2}$ 
(see Figure \ref{figure:validity}),
no data is available at those scales. 
At the energy scale { $Q^{2}=0.55\,{\rm GeV^{2}}$},
x is restricted to the range
\begin{equation}
0.035<x<0.060.
\end{equation}
Figure
\ref{figure:validity} shows the approximate 
validity region of the 3dMVn model
for a gold nucleus. We consider gold because 
it provides for the
greatest range of validity of the model as 
represented by Eq. (\ref{lowerlim}).
The results for lead would be indistinguishable. 

The parameters of the 3dMVn model need to be fixed 
before performing the numerical integration
in Eq. (\ref{eq:gluon-distro-final}). We take the 
nucleon mass $m=0.94\,\mbox{GeV}$,
$R_0=1\,{\rm fm}$ and $A=197$.
The coupling constant $\alpha_{s}$ and the 
correlation length $\lambda$ 
are treated as free parameters.
We evaluate $xg_{A}(x,Q^{2})$ numerically at 
$Q^2=0.55\,\rm{GeV^2}$ in the range
$0.035<x<0.060$ for several values of 
$\alpha_{s}$ and $\lambda$
in increments of 0.1
in the range 
\begin{equation}
 \begin{aligned}
0.1 \leq &\alpha_{s}\leq  1,\\
0.2\,{\rm fm}  \leq & \lambda \leq  
2.6\,{\rm fm.}\label{eq:param}
  \end{aligned}
\end{equation}
Then we fit $xg_{A}(x,Q^{2})$ to a form
\begin{equation}
xg_{A}(x,Q^{2})= b \, x^c \, (1-x)^d,
\label{parametrization}
\end{equation}
and find $\{b,c,d\}$ for each set of $\{\alpha_s,\lambda\}$.
After the numerical integration, we obtain 
${xg_{197}(x,Q^{2})}$
for the whole nucleus
for several points in the parameter space 
$\{\alpha_{S},\lambda \}$.  
The scale $Q^{2}=0.55\,{\rm GeV^{2}}$ at 
which we calculated $xg_{A}(x,Q^{2})$
is the lower limit of the JR09 data 
parametrization for the parton
distribution functions.
For this reason, we compare JR09 with the gluon distribution functions that are calculated and evolved to $Q^{2}=100\,{\rm GeV^{2}}$.

Gluon distribution functions are evolved to different energy scales
with the DGLAP equation 
\cite{Gribov:1972ri,Lipatov:1974qm,Dokshitzer:1977sg,Altarelli:1977zs}. 
For this purpose, we employ the code 
QCDNUM17 \cite{Botje:2010ay}.
We will assume that nucleus is a dilute system of nucleons
and take the nuclear modification factor $R_A=1$ 
\cite{JalilianMarian:1999cf}. Hence,
\begin{equation}
xg_{p/A}(x,Q^{2})\equiv\frac{1}{A}xg_{A}(x,Q^{2}),
\end{equation}
where the subscript $p/A$ refers to a nucleon in 
a nuclear environment. 
Henceforth, we will use the gluon distribution function for
a single nucleon in a gold nucleus.
It should be kept in mind that 
$g_{p/A}(x,Q^{2})$ is in principle different than a
gluon distribution function for an isolated
proton $g_{p}(x,Q^{2})$ since the nuclear effects
introduce enhancement or shadowing 
depending on the energy scale $Q^{2}$.

We run QCDNUM in the variable-flavor number scheme (VFNS)  
and at the next-to-next-to-leading order (NNLO($\overline{\text{MS}}$)). 
For the DGLAP evolution of the gluons, valence and 
sea quark distributions
at the initial scale should also be provided to the evolution
code. At the initial scale $Q^{2}=0.55\,{\rm GeV^{2}}$, the calculated
gluon distribution function is the main input to QCDNUM. As for
the valence and sea quark distributions, the JR09 
\cite{JimenezDelgado:2009tv} data
parametrization is utilized. The evolution is 
repeated for several
values of $\alpha_{s}$ and $\lambda$ 
in the range given in Eq. (\ref{eq:param}).


\section{Data Parametrization}

We will compare our results with the parametrization of data by 
JR09 \cite{JimenezDelgado:2009tv}.
We use the PDFs that work in the VFNS scheme and at the
NNLO($\overline{\text{MS}}$) order.
The JR09 is valid in the ranges
\begin{equation}
 \begin{aligned}
0.55\lesssim & Q^{2}\lesssim10^{8}\,{\rm GeV^{2}}, \\
10^{-9}\lesssim & x\lesssim1.
  \end{aligned}
\end{equation}

The JR09 provides
PDFs for separate nucleons $xg_{p}(x,Q^{2})$ but
it does not contain any information about the nuclear modification
function $R_A$. Our results, on the other hand, are for a nucleon in
a nuclear environment $xg_{p/A}(x,Q^{2})$; therefore, 
$R_A$ should be, in principle, taken into account. However, 
the nuclear PDFs (for example nCTEQ 
\cite{Schienbein:2009kk}) are available only for
$Q^{2}>1\,{\rm GeV^{2}}$.
As pointed out earlier, since the DGLAP equation mixes quarks and
gluons during the evolution, we need the valence and sea quark distribution
functions at the initial energy scale so that the gluon distribution function 
calculated from the model can be evolved to the higher energies. 
Hence, for consistency, we will utilize
the quark PDFs provided by JR09 at $Q^{2}=0.55\,{\rm GeV^{2}}$ as an
input to the DGLAP evolution. 
The discrepancy
between JR09 and nCTEQ \cite{Schienbein:2009kk},
encoded in the nuclear modification factor,
is at most $5\%$ at $Q^{2}=25\,{\rm GeV^{2}}$. 
At $Q^{2}=100$ and $1000\,{\rm GeV^{2}}$ the discrepancy is
negligibly small. For that reason, we think 
the lack of information regarding
$R_A$ will not cause a significant error in our analysis.


\section{Best Fit Parameters}
Now we seek the sets of parameters $\{\alpha_s,\lambda \}$
that produce the best fits. For this reason,
we compare the gluon distribution functions, 
calculated and evolved to $Q^{2}=100\,{\rm GeV^{2}}$, 
with JR09 at the same energy.

In the DGLAP evolution to higher energies, the contribution 
to the radiation at a particular value of x always 
comes from the sources from the 
larger x region.
Hence, the effect of gluons calculated from 3dMVn 
in the range $0.035<x<0.060$ at 
$Q^{2}=0.55\,{\rm GeV^{2}}$ will be more prominent 
at smaller x values at
$Q^{2}=100\,{\rm GeV^{2}}$. For this purpose,
we will compare the evolved model with JR09 in the range
\begin{equation}
0.015<x<0.04.
\label{fitregion}
\end{equation}
The gluons from $x<0.015$ do not enter the DGLAP evolution
and hence they do not contribute to the region in Eq. (\ref{fitregion}).
The gluons from $0.015<x<1$ will enter the 
DGLAP evolution. We argue that the
parametrization in Eq. (\ref{parametrization}) is a 
good approximation
for the large-x gluons.

\begin{figure}[t]
\begin{centering}
\includegraphics[scale=0.7]{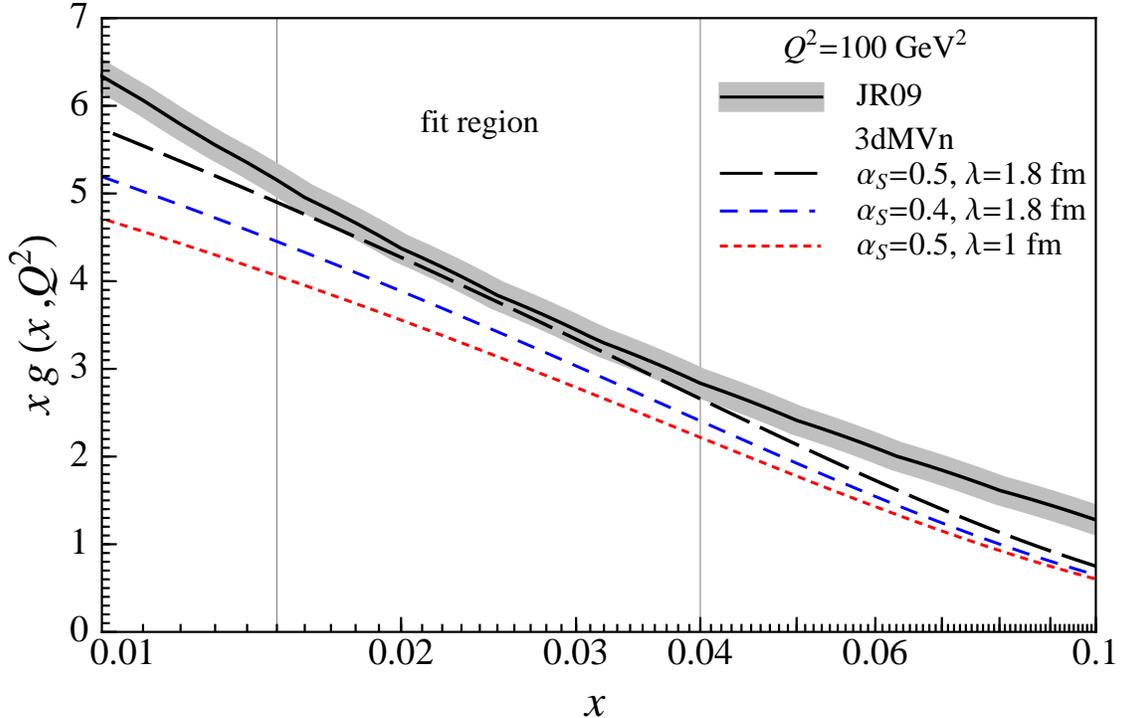}
\par\end{centering} 
\caption{Comparison of the gluon distribution
function of a nucleon from the 3dMVn model 
for various values of $\{\alpha_s,\lambda\}$ 
with the JR09 data parametrization at 
$Q^2= 100\,\rm{GeV^2}$. The horizontal axis is
in logarithmic scale.
For distances larger than the correlation
length $\lambda$, gluons are not
correlated and hence the color neutrality condition
is satisfied.
The model is reliable in the
fit region bounded by the two vertical lines. 
The uncertainty in JR09 is shown with 
an error band and it is about $5\%$.
The discrepancy between the best fit curve 
($\alpha_s=0.5$ and $\lambda=1.8\,\rm{fm}$)
and the JR09 in the fit region is only $2\%$. 
}
\label{fig:modelanddata}
\end{figure}

Our analysis shows that the best fit occurs at the values 
$\alpha_{s}=0.5$ and
$\lambda=1.8\,{\rm fm}$. In Figure \ref{fig:modelanddata} 
we show a comparison between the JR09 data parametrization
and 3dMVn model at $Q^{2}=100\,{\rm GeV^{2}}$ for 
$\alpha_{s}=0.5$ and
$\lambda=1.8\,{\rm fm}$.
We find almost identical plots at other $Q^2$ 
values in the range
${0.015<x<0.04}$.

At other values of $\alpha_s$ and $\lambda$, 
we find that the model underestimates the data. 
As $\alpha_s$ increases, the 3dMVn curve increases
and the disagreement between the model and JR09 decreases. 
For values $\alpha_s \geq 0.5$ 
the model does not change significantly. 
Hence, we take
${\alpha_s=0.5}$ since this is the smallest value of $\alpha_s$ 
for which a good fit can be obtained.
Interestingly, this ``freezing'' behavior of the 
strong coupling constant
particularly at $\alpha_s=0.5$,
imposed in other approaches by hand 
(see \cite{Dumitru:2011wq,Albacete:2012xq}),
arises naturally in the 3dMVn model.
Adjusting $\lambda$ changes the model curve only slightly.
For $\alpha_s=0.4$ and 
values of $\lambda$ greater than $1.8\,\rm{fm}$, 
we do not find
a fit as good as the one for $\alpha_s=0.5$ 
and $\lambda=1.8\,\rm{fm}$.

There are various sources of uncertainty 
in the MV model which are shared with the 3dMVn model. 
In the paradigm of strong classical color fields, 
one wishes to employ solutions of
the classical Yang-Mills equation which is given 
in covariant gauge by
\begin{equation}
(\boldsymbol{\nabla}_{\perp}^2+ \partial_{\parallel}^2)A^\nu= 
g J^\nu
+
2ig[A^\mu,\partial_\mu A^\nu]
-ig[A_\mu,\partial^\nu A^\mu]
+
g^2[A_\mu,[A^\mu,A^\nu]]
,
\label{cym}
\end{equation}
where $\partial_{\parallel}^2=-2 \partial_{+} \partial_{-}$. 
The second and third terms 
on the right hand side of the Eq. (\ref{cym}) are 
responsible for the
three gluon interaction vertex, and the last term
is for the four gluon interaction vertex in the quantized
theory. In the framework of Color Glass Condensate,
only the linearized version of Eq. (\ref{cym}) is used.
The nonabelian feature of the model sets in when 
the solutions of the linearized equation in the covariant
gauge are transformed to the light-cone gauge by
means of the full gauge transformation including the
nonabelian term \cite{Lam:1999wu}
\begin{equation}
A_{\rm{LC}}^{\mu}(x)={\rm{U}}(x)  
A_{\rm{cov}}^{\mu} {\rm{U}}^{-1}(x) 
- \frac{i}{g} [\partial^{\mu}  {\rm{U}}(x)] {\rm{U}}^{-1}(x). 
\end{equation}
The reason why the nonlinear terms in Eq. (\ref{cym}) 
are omitted
is simply because the solutions of the fully nonlinear theory
are not known. In addition, the Green's function for gluons,
which is essential in the calculation of 
{$\langle A_{i}^{a}(\boldsymbol{x})A_{i}^{b}(
\boldsymbol{x}^{\,\prime})\rangle$} from 
{$\langle\rho^{a}(\boldsymbol{x})\rho^{b}(
\boldsymbol{x}^{\,\prime})\rangle$},
can only be defined in the linear theory. 
The gluon propagator is calculated from
\begin{equation}
(\boldsymbol{\nabla}_{\perp}^2+ \partial_{\parallel}^2) 
G(\boldsymbol{x})= \delta^3 (\boldsymbol{x}).
\end{equation}

The origin of the infrared divergences in the MV model
lies here and this is also why confinement effects need to be
introduced by hand by using colored noise. In principle,
the correlation function with colored noise
in Eq. (\ref{eq:Yukawa})
should be calculable from the fully nonlinear theory.

The linear
approximation can only be justified if $g \ll 1$
so that
the nonlinear terms can be neglected.
However, 
the strong coupling constant is not expected to
be so small and nonlinear terms are as 
important as the source term.
Hence, the linear approximation
of Eq. (\ref{cym}) is the main source of the uncertainty 
in any realistic analysis.

Another source of uncertainty 
may be the assumption that 
the nucleus is taken to be much larger
than a nucleon, $R \gg R_0$. For a gold
nucleus, the corrections may be as large as $17\%$ since
$R_0/R \sim A^{-1/3} \sim 0.17. $
Also quantum corrections at 
smaller values of x may be important
even in the weak coupling limit. These corrections
are of the form $\alpha_s \ln (1/x)$ and
discussed in the references
 \cite{Ayala:1995kg,Ayala:1995hx,Lam:2000nz}. 
Lastly, although we estimate that the error due to
neglecting the nuclear effects should be small,
it would be worth of repeating the analysis
presented in this paper by using nuclear PDFs
once they are available for $Q^2 \sim 0.55\, \rm{GeV^2}$.


\section{Summary and Outlook}

The framework of Color Glass Condensate allows 
one to calculate the vector field correlation
function. It can be used to calculate the gluon 
distribution function of the ultrarelativistic nuclei and
the initial energy density distribution due to the 
interacting classical color fields
produced by the colliding nuclei in heavy ion collisions.

We have examined a three dimensional color neutral version of
the McLerran-Venugopalan model. The 3dMVn 
model is finite in the infrared
and therefore an infrared cutoff is not needed.
In addition, the results of this model are
x dependent due to the intrinsic three dimensional 
treatment of the nucleus
in contrast to the approximation of infinitely thin nucleus.

In order to explore the parameter space of these 
two variables, we have calculated the gluon distribution 
function for several values of $\alpha_s$ and $\lambda$.
The originality of this work is to compare our calculations directly with parametrization of the data to determine the free parameters of the model. We have found the best fit between
the gluon distribution function from the JR09 parametrization
and the one calculated from the 3dMVn model occurs at
$\alpha_s=0.5$ and $\lambda=1.8\,\rm{fm}.$
At other values of these two parameters the model underestimates
the data. This may be due to the uncertainty 
in the assumptions and 
rough estimates made during the construction of the model
as well as the other uncertainties discussed in the previous
section. 
We have also found that the 3dMVn model had an intrinsic freezing behavior that the gluon distribution function froze at $\alpha_s=0.5$ and remained unchanged for $\alpha_s>0.5$.

The assumption that the color charge is normally distributed throughout the nucleus lies at the heart of the MV and 3dMVn models. Besides the normal (Gaussian) distribution of the color charge, the nucleonic inner structure of the nucleus is implemented via colored noise by introducing correlations for distances smaller than a nucleon size. In this manner, the fluctuations in the positions of nucleons in a nucleus have not been treated separately from the dynamical color charge fluctuations. In other words, the effect of confinement is realized through short range correlations in $\langle\rho^{a}(\boldsymbol{x})\rho^{b}(\boldsymbol{x}^{\,\prime})\rangle$ by colored noise rather than considering the nucleus as a collection of individual nucleons. An alternative to this might be an event-by-event Monte Carlo sampling of the distribution of nucleons \cite{Schenke:2010rr,Schenke:2011tv,Schenke:2012wb,Schenke:2012hg}. For a finite nucleus, the results from event-by-event fluctuations in positions of the sampled nucleons may differ from the analytical calculations presented in this paper. For a very large nucleus, we expect the results from both approaches to agree.

The final product of this work is the unintegrated gluon distribution (also known as the ``gluon number density'') given in Eq. (\ref{eq:gluon-distro-LM}) of which parameters are fixed in the previous section. This three dimensional quantity is the main ingredient for various applications of the CGC concept. It can be picked up by the practitioners of CGC and be used immediately in realistic calculations other than calculation of the gluon distribution function.

In this work, we calculated the gluon distribution function since it was a quantity that could be easily compared with data.
However, 
the ultimate goal was not 
calculating the gluon distribution function per se, 
but to attain a working model for further use, 
particularly in the calculation of the initial energy density
distribution in heavy ion collisions.
This work is in progress.
In the future work, we plan to calculate the initial 
energy density in heavy ion collision in the framework of the 3dMVn model.

\section*{Acknowledgments}
We thank Joseph Kapusta and Rainer J. Fries for 
useful discussions and their critical comments on the manuscript.
We also thank Michiel Botje for correspondence regarding
his QCDNUM code. This work was supported by the U.S. DOE Award
No. DE-FG02-87ER40328.

\raggedright
\bibliography{3dMVn-vs-JR09}

\begin{thebibliography}{24}%
\makeatletter
\providecommand \@ifxundefined [1]{%
 \@ifx{#1\undefined}
}%
\providecommand \@ifnum [1]{%
 \ifnum #1\expandafter \@firstoftwo
 \else \expandafter \@secondoftwo
 \fi
}%
\providecommand \@ifx [1]{%
 \ifx #1\expandafter \@firstoftwo
 \else \expandafter \@secondoftwo
 \fi
}%
\providecommand \natexlab [1]{#1}%
\providecommand \enquote  [1]{``#1''}%
\providecommand \bibnamefont  [1]{#1}%
\providecommand \bibfnamefont [1]{#1}%
\providecommand \citenamefont [1]{#1}%
\providecommand \href@noop [0]{\@secondoftwo}%
\providecommand \href [0]{\begingroup \@sanitize@url \@href}%
\providecommand \@href[1]{\@@startlink{#1}\@@href}%
\providecommand \@@href[1]{\endgroup#1\@@endlink}%
\providecommand \@sanitize@url [0]{\catcode `\\12\catcode `\$12\catcode
  `\&12\catcode `\#12\catcode `\^12\catcode `\_12\catcode `\%12\relax}%
\providecommand \@@startlink[1]{}%
\providecommand \@@endlink[0]{}%
\providecommand \url  [0]{\begingroup\@sanitize@url \@url }%
\providecommand \@url [1]{\endgroup\@href {#1}{\urlprefix }}%
\providecommand \urlprefix  [0]{URL }%
\providecommand \Eprint [0]{\href }%
\providecommand \doibase [0]{http://dx.doi.org/}%
\providecommand \selectlanguage [0]{\@gobble}%
\providecommand \bibinfo  [0]{\@secondoftwo}%
\providecommand \bibfield  [0]{\@secondoftwo}%
\providecommand \translation [1]{[#1]}%
\providecommand \BibitemOpen [0]{}%
\providecommand \bibitemStop [0]{}%
\providecommand \bibitemNoStop [0]{.\EOS\space}%
\providecommand \EOS [0]{\spacefactor3000\relax}%
\providecommand \BibitemShut  [1]{\csname bibitem#1\endcsname}%
\let\auto@bib@innerbib\@empty
\bibitem [{\citenamefont {Iancu}\ and\ \citenamefont
  {Venugopalan}(2003)}]{Iancu:2003xm}%
  \BibitemOpen
  \bibfield  {author} {\bibinfo {author} {\bibfnamefont {E.}~\bibnamefont
  {Iancu}}\ and\ \bibinfo {author} {\bibfnamefont {R.}~\bibnamefont
  {Venugopalan}},\ }\href@noop {} {\bibfield  {journal} {\bibinfo  {journal}
  {In *Hwa, R.C. (ed.) et al.: Quark Gluon Plasma*}\ ,\ \bibinfo {pages} {249}}
  (\bibinfo {year} {2003})},\ \Eprint {http://arxiv.org/abs/hep-ph/0303204}
  {arXiv:hep-ph/0303204 [hep-ph]} \BibitemShut {NoStop}%
\bibitem [{\citenamefont {McLerran}\ and\ \citenamefont
  {Venugopalan}(1994{\natexlab{a}})}]{McLerran:1993ni}%
  \BibitemOpen
  \bibfield  {author} {\bibinfo {author} {\bibfnamefont {L.~D.}\ \bibnamefont
  {McLerran}}\ and\ \bibinfo {author} {\bibfnamefont {R.}~\bibnamefont
  {Venugopalan}},\ }\href {\doibase 10.1103/PhysRevD.49.2233} {\bibfield
  {journal} {\bibinfo  {journal} {Phys.Rev.}\ }\textbf {\bibinfo {volume}
  {D49}},\ \bibinfo {pages} {2233} (\bibinfo {year} {1994}{\natexlab{a}})},\
  \Eprint {http://arxiv.org/abs/hep-ph/9309289} {arXiv:hep-ph/9309289 [hep-ph]}
  \BibitemShut {NoStop}%
\bibitem [{\citenamefont {McLerran}\ and\ \citenamefont
  {Venugopalan}(1994{\natexlab{b}})}]{McLerran:1993ka}%
  \BibitemOpen
  \bibfield  {author} {\bibinfo {author} {\bibfnamefont {L.~D.}\ \bibnamefont
  {McLerran}}\ and\ \bibinfo {author} {\bibfnamefont {R.}~\bibnamefont
  {Venugopalan}},\ }\href {\doibase 10.1103/PhysRevD.49.3352} {\bibfield
  {journal} {\bibinfo  {journal} {Phys.Rev.}\ }\textbf {\bibinfo {volume}
  {D49}},\ \bibinfo {pages} {3352} (\bibinfo {year} {1994}{\natexlab{b}})},\
  \Eprint {http://arxiv.org/abs/hep-ph/9311205} {arXiv:hep-ph/9311205 [hep-ph]}
  \BibitemShut {NoStop}%
\bibitem [{\citenamefont {McLerran}\ and\ \citenamefont
  {Venugopalan}(1994{\natexlab{c}})}]{McLerran:1994vd}%
  \BibitemOpen
  \bibfield  {author} {\bibinfo {author} {\bibfnamefont {L.~D.}\ \bibnamefont
  {McLerran}}\ and\ \bibinfo {author} {\bibfnamefont {R.}~\bibnamefont
  {Venugopalan}},\ }\href {\doibase 10.1103/PhysRevD.50.2225} {\bibfield
  {journal} {\bibinfo  {journal} {Phys.Rev.}\ }\textbf {\bibinfo {volume}
  {D50}},\ \bibinfo {pages} {2225} (\bibinfo {year} {1994}{\natexlab{c}})},\
  \Eprint {http://arxiv.org/abs/hep-ph/9402335} {arXiv:hep-ph/9402335 [hep-ph]}
  \BibitemShut {NoStop}%
\bibitem [{\citenamefont {Ayala}\ \emph {et~al.}(1995)\citenamefont {Ayala},
  \citenamefont {Jalilian-Marian}, \citenamefont {McLerran},\ and\
  \citenamefont {Venugopalan}}]{Ayala:1995kg}%
  \BibitemOpen
  \bibfield  {author} {\bibinfo {author} {\bibfnamefont {A.}~\bibnamefont
  {Ayala}}, \bibinfo {author} {\bibfnamefont {J.}~\bibnamefont
  {Jalilian-Marian}}, \bibinfo {author} {\bibfnamefont {L.~D.}\ \bibnamefont
  {McLerran}}, \ and\ \bibinfo {author} {\bibfnamefont {R.}~\bibnamefont
  {Venugopalan}},\ }\href {\doibase 10.1103/PhysRevD.52.2935} {\bibfield
  {journal} {\bibinfo  {journal} {Phys.Rev.}\ }\textbf {\bibinfo {volume}
  {D52}},\ \bibinfo {pages} {2935} (\bibinfo {year} {1995})},\ \Eprint
  {http://arxiv.org/abs/hep-ph/9501324} {arXiv:hep-ph/9501324 [hep-ph]}
  \BibitemShut {NoStop}%
\bibitem [{\citenamefont {Ayala}\ \emph {et~al.}(1996)\citenamefont {Ayala},
  \citenamefont {Jalilian-Marian}, \citenamefont {McLerran},\ and\
  \citenamefont {Venugopalan}}]{Ayala:1995hx}%
  \BibitemOpen
  \bibfield  {author} {\bibinfo {author} {\bibfnamefont {A.}~\bibnamefont
  {Ayala}}, \bibinfo {author} {\bibfnamefont {J.}~\bibnamefont
  {Jalilian-Marian}}, \bibinfo {author} {\bibfnamefont {L.~D.}\ \bibnamefont
  {McLerran}}, \ and\ \bibinfo {author} {\bibfnamefont {R.}~\bibnamefont
  {Venugopalan}},\ }\href {\doibase 10.1103/PhysRevD.53.458} {\bibfield
  {journal} {\bibinfo  {journal} {Phys.Rev.}\ }\textbf {\bibinfo {volume}
  {D53}},\ \bibinfo {pages} {458} (\bibinfo {year} {1996})},\ \Eprint
  {http://arxiv.org/abs/hep-ph/9508302} {arXiv:hep-ph/9508302 [hep-ph]}
  \BibitemShut {NoStop}%
\bibitem [{\citenamefont {Lam}\ and\ \citenamefont
  {Mahlon}(1999)}]{Lam:1999wu}%
  \BibitemOpen
  \bibfield  {author} {\bibinfo {author} {\bibfnamefont {C.}~\bibnamefont
  {Lam}}\ and\ \bibinfo {author} {\bibfnamefont {G.}~\bibnamefont {Mahlon}},\
  }\href {\doibase 10.1103/PhysRevD.61.014005} {\bibfield  {journal} {\bibinfo
  {journal} {Phys.Rev.}\ }\textbf {\bibinfo {volume} {D61}},\ \bibinfo {pages}
  {014005} (\bibinfo {year} {1999})},\ \Eprint
  {http://arxiv.org/abs/hep-ph/9907281} {arXiv:hep-ph/9907281 [hep-ph]}
  \BibitemShut {NoStop}%
\bibitem [{\citenamefont {Lam}\ and\ \citenamefont
  {Mahlon}(2000)}]{Lam:2000nz}%
  \BibitemOpen
  \bibfield  {author} {\bibinfo {author} {\bibfnamefont {C.}~\bibnamefont
  {Lam}}\ and\ \bibinfo {author} {\bibfnamefont {G.}~\bibnamefont {Mahlon}},\
  }\href {\doibase 10.1103/PhysRevD.62.114023} {\bibfield  {journal} {\bibinfo
  {journal} {Phys.Rev.}\ }\textbf {\bibinfo {volume} {D62}},\ \bibinfo {pages}
  {114023} (\bibinfo {year} {2000})},\ \Eprint
  {http://arxiv.org/abs/hep-ph/0007133} {arXiv:hep-ph/0007133 [hep-ph]}
  \BibitemShut {NoStop}%
\bibitem [{\citenamefont {Jalilian-Marian}\ \emph {et~al.}(1997)\citenamefont
  {Jalilian-Marian}, \citenamefont {Kovner}, \citenamefont {McLerran},\ and\
  \citenamefont {Weigert}}]{JalilianMarian:1996xn}%
  \BibitemOpen
  \bibfield  {author} {\bibinfo {author} {\bibfnamefont {J.}~\bibnamefont
  {Jalilian-Marian}}, \bibinfo {author} {\bibfnamefont {A.}~\bibnamefont
  {Kovner}}, \bibinfo {author} {\bibfnamefont {L.~D.}\ \bibnamefont
  {McLerran}}, \ and\ \bibinfo {author} {\bibfnamefont {H.}~\bibnamefont
  {Weigert}},\ }\href {\doibase 10.1103/PhysRevD.55.5414} {\bibfield  {journal}
  {\bibinfo  {journal} {Phys.Rev.}\ }\textbf {\bibinfo {volume} {D55}},\
  \bibinfo {pages} {5414} (\bibinfo {year} {1997})},\ \Eprint
  {http://arxiv.org/abs/hep-ph/9606337} {arXiv:hep-ph/9606337 [hep-ph]}
  \BibitemShut {NoStop}%
\bibitem [{\citenamefont {Kovchegov}(1996)}]{Kovchegov:1996ty}%
  \BibitemOpen
  \bibfield  {author} {\bibinfo {author} {\bibfnamefont {Y.~V.}\ \bibnamefont
  {Kovchegov}},\ }\href {\doibase 10.1103/PhysRevD.54.5463} {\bibfield
  {journal} {\bibinfo  {journal} {Phys.Rev.}\ }\textbf {\bibinfo {volume}
  {D54}},\ \bibinfo {pages} {5463} (\bibinfo {year} {1996})},\ \Eprint
  {http://arxiv.org/abs/hep-ph/9605446} {arXiv:hep-ph/9605446 [hep-ph]}
  \BibitemShut {NoStop}%
\bibitem [{\citenamefont {Gribov}\ and\ \citenamefont
  {Lipatov}(1972)}]{Gribov:1972ri}%
  \BibitemOpen
  \bibfield  {author} {\bibinfo {author} {\bibfnamefont {V.}~\bibnamefont
  {Gribov}}\ and\ \bibinfo {author} {\bibfnamefont {L.}~\bibnamefont
  {Lipatov}},\ }\href@noop {} {\bibfield  {journal} {\bibinfo  {journal}
  {Sov.J.Nucl.Phys.}\ }\textbf {\bibinfo {volume} {15}},\ \bibinfo {pages}
  {438} (\bibinfo {year} {1972})}\BibitemShut {NoStop}%
\bibitem [{\citenamefont {Lipatov}(1975)}]{Lipatov:1974qm}%
  \BibitemOpen
  \bibfield  {author} {\bibinfo {author} {\bibfnamefont {L.}~\bibnamefont
  {Lipatov}},\ }\href@noop {} {\bibfield  {journal} {\bibinfo  {journal}
  {Sov.J.Nucl.Phys.}\ }\textbf {\bibinfo {volume} {20}},\ \bibinfo {pages} {94}
  (\bibinfo {year} {1975})}\BibitemShut {NoStop}%
\bibitem [{\citenamefont {Dokshitzer}(1977)}]{Dokshitzer:1977sg}%
  \BibitemOpen
  \bibfield  {author} {\bibinfo {author} {\bibfnamefont {Y.~L.}\ \bibnamefont
  {Dokshitzer}},\ }\href@noop {} {\bibfield  {journal} {\bibinfo  {journal}
  {Sov.Phys.JETP}\ }\textbf {\bibinfo {volume} {46}},\ \bibinfo {pages} {641}
  (\bibinfo {year} {1977})}\BibitemShut {NoStop}%
\bibitem [{\citenamefont {Altarelli}\ and\ \citenamefont
  {Parisi}(1977)}]{Altarelli:1977zs}%
  \BibitemOpen
  \bibfield  {author} {\bibinfo {author} {\bibfnamefont {G.}~\bibnamefont
  {Altarelli}}\ and\ \bibinfo {author} {\bibfnamefont {G.}~\bibnamefont
  {Parisi}},\ }\href {\doibase 10.1016/0550-3213(77)90384-4} {\bibfield
  {journal} {\bibinfo  {journal} {Nucl.Phys.}\ }\textbf {\bibinfo {volume}
  {B126}},\ \bibinfo {pages} {298} (\bibinfo {year} {1977})}\BibitemShut
  {NoStop}%
\bibitem [{\citenamefont {Botje}(2011)}]{Botje:2010ay}%
  \BibitemOpen
  \bibfield  {author} {\bibinfo {author} {\bibfnamefont {M.}~\bibnamefont
  {Botje}},\ }\href {\doibase 10.1016/j.cpc.2010.10.020} {\bibfield  {journal}
  {\bibinfo  {journal} {Comput.Phys.Commun.}\ }\textbf {\bibinfo {volume}
  {182}},\ \bibinfo {pages} {490} (\bibinfo {year} {2011})},\ \Eprint
  {http://arxiv.org/abs/1005.1481} {arXiv:1005.1481 [hep-ph]} \BibitemShut
  {NoStop}%
\bibitem [{\citenamefont {Jalilian-Marian}\ and\ \citenamefont
  {Wang}(1999)}]{JalilianMarian:1999cf}%
  \BibitemOpen
  \bibfield  {author} {\bibinfo {author} {\bibfnamefont {J.}~\bibnamefont
  {Jalilian-Marian}}\ and\ \bibinfo {author} {\bibfnamefont {X.-N.}\
  \bibnamefont {Wang}},\ }\href {\doibase 10.1103/PhysRevD.60.054016}
  {\bibfield  {journal} {\bibinfo  {journal} {Phys.Rev.}\ }\textbf {\bibinfo
  {volume} {D60}},\ \bibinfo {pages} {054016} (\bibinfo {year} {1999})},\
  \Eprint {http://arxiv.org/abs/hep-ph/9902411} {arXiv:hep-ph/9902411 [hep-ph]}
  \BibitemShut {NoStop}%
\bibitem [{\citenamefont {Jimenez-Delgado}\ and\ \citenamefont
  {Reya}(2009)}]{JimenezDelgado:2009tv}%
  \BibitemOpen
  \bibfield  {author} {\bibinfo {author} {\bibfnamefont {P.}~\bibnamefont
  {Jimenez-Delgado}}\ and\ \bibinfo {author} {\bibfnamefont {E.}~\bibnamefont
  {Reya}},\ }\href {\doibase 10.1103/PhysRevD.80.114011} {\bibfield  {journal}
  {\bibinfo  {journal} {Phys.Rev.}\ }\textbf {\bibinfo {volume} {D80}},\
  \bibinfo {pages} {114011} (\bibinfo {year} {2009})},\ \Eprint
  {http://arxiv.org/abs/0909.1711} {arXiv:0909.1711 [hep-ph]} \BibitemShut
  {NoStop}%
\bibitem [{\citenamefont {Schienbein}\ \emph {et~al.}(2009)\citenamefont
  {Schienbein}, \citenamefont {Yu}, \citenamefont {Kovarik}, \citenamefont
  {Keppel}, \citenamefont {Morfin}, \citenamefont {Olness},\ and\ \citenamefont
  {Owens}}]{Schienbein:2009kk}%
  \BibitemOpen
  \bibfield  {author} {\bibinfo {author} {\bibfnamefont {I.}~\bibnamefont
  {Schienbein}}, \bibinfo {author} {\bibfnamefont {J.}~\bibnamefont {Yu}},
  \bibinfo {author} {\bibfnamefont {K.}~\bibnamefont {Kovarik}}, \bibinfo
  {author} {\bibfnamefont {C.}~\bibnamefont {Keppel}}, \bibinfo {author}
  {\bibfnamefont {J.}~\bibnamefont {Morfin}}, \bibinfo {author} {\bibfnamefont
  {F.}~\bibnamefont {Olness}}, \ and\ \bibinfo {author} {\bibfnamefont
  {J.}~\bibnamefont {Owens}},\ }\href {\doibase 10.1103/PhysRevD.80.094004}
  {\bibfield  {journal} {\bibinfo  {journal} {Phys.Rev.}\ }\textbf {\bibinfo
  {volume} {D80}},\ \bibinfo {pages} {094004} (\bibinfo {year} {2009})},\
  \Eprint {http://arxiv.org/abs/0907.2357} {arXiv:0907.2357 [hep-ph]}
  \BibitemShut {NoStop}%
\bibitem [{\citenamefont {Dumitru}\ \emph {et~al.}(2012)\citenamefont
  {Dumitru}, \citenamefont {Kharzeev}, \citenamefont {Levin},\ and\
  \citenamefont {Nara}}]{Dumitru:2011wq}%
  \BibitemOpen
  \bibfield  {author} {\bibinfo {author} {\bibfnamefont {A.}~\bibnamefont
  {Dumitru}}, \bibinfo {author} {\bibfnamefont {D.~E.}\ \bibnamefont
  {Kharzeev}}, \bibinfo {author} {\bibfnamefont {E.~M.}\ \bibnamefont {Levin}},
  \ and\ \bibinfo {author} {\bibfnamefont {Y.}~\bibnamefont {Nara}},\ }\href
  {\doibase 10.1103/PhysRevC.85.044920} {\bibfield  {journal} {\bibinfo
  {journal} {Phys.Rev.}\ }\textbf {\bibinfo {volume} {C85}},\ \bibinfo {pages}
  {044920} (\bibinfo {year} {2012})},\ \Eprint {http://arxiv.org/abs/1111.3031}
  {arXiv:1111.3031 [hep-ph]} \BibitemShut {NoStop}%
\bibitem [{\citenamefont {Albacete}\ \emph {et~al.}(2012)\citenamefont
  {Albacete}, \citenamefont {Dumitru}, \citenamefont {Fujii},\ and\
  \citenamefont {Nara}}]{Albacete:2012xq}%
  \BibitemOpen
  \bibfield  {author} {\bibinfo {author} {\bibfnamefont {J.~L.}\ \bibnamefont
  {Albacete}}, \bibinfo {author} {\bibfnamefont {A.}~\bibnamefont {Dumitru}},
  \bibinfo {author} {\bibfnamefont {H.}~\bibnamefont {Fujii}}, \ and\ \bibinfo
  {author} {\bibfnamefont {Y.}~\bibnamefont {Nara}},\ }\href@noop {} {\
  (\bibinfo {year} {2012})},\ \Eprint {http://arxiv.org/abs/1209.2001}
  {arXiv:1209.2001 [hep-ph]} \BibitemShut {NoStop}%
\bibitem [{\citenamefont {Schenke}\ \emph
  {et~al.}(2011{\natexlab{a}})\citenamefont {Schenke}, \citenamefont {Jeon},\
  and\ \citenamefont {Gale}}]{Schenke:2010rr}%
  \BibitemOpen
  \bibfield  {author} {\bibinfo {author} {\bibfnamefont {B.}~\bibnamefont
  {Schenke}}, \bibinfo {author} {\bibfnamefont {S.}~\bibnamefont {Jeon}}, \
  and\ \bibinfo {author} {\bibfnamefont {C.}~\bibnamefont {Gale}},\ }\href
  {\doibase 10.1103/PhysRevLett.106.042301} {\bibfield  {journal} {\bibinfo
  {journal} {Phys.Rev.Lett.}\ }\textbf {\bibinfo {volume} {106}},\ \bibinfo
  {pages} {042301} (\bibinfo {year} {2011}{\natexlab{a}})},\ \Eprint
  {http://arxiv.org/abs/1009.3244} {arXiv:1009.3244 [hep-ph]} \BibitemShut
  {NoStop}%
\bibitem [{\citenamefont {Schenke}\ \emph
  {et~al.}(2011{\natexlab{b}})\citenamefont {Schenke}, \citenamefont {Jeon},\
  and\ \citenamefont {Gale}}]{Schenke:2011tv}%
  \BibitemOpen
  \bibfield  {author} {\bibinfo {author} {\bibfnamefont {B.}~\bibnamefont
  {Schenke}}, \bibinfo {author} {\bibfnamefont {S.}~\bibnamefont {Jeon}}, \
  and\ \bibinfo {author} {\bibfnamefont {C.}~\bibnamefont {Gale}},\ }\href
  {\doibase 10.1016/j.physletb.2011.06.065} {\bibfield  {journal} {\bibinfo
  {journal} {Phys.Lett.}\ }\textbf {\bibinfo {volume} {B702}},\ \bibinfo
  {pages} {59} (\bibinfo {year} {2011}{\natexlab{b}})},\ \Eprint
  {http://arxiv.org/abs/1102.0575} {arXiv:1102.0575 [hep-ph]} \BibitemShut
  {NoStop}%
\bibitem [{\citenamefont {Schenke}\ \emph
  {et~al.}(2012{\natexlab{a}})\citenamefont {Schenke}, \citenamefont
  {Tribedy},\ and\ \citenamefont {Venugopalan}}]{Schenke:2012wb}%
  \BibitemOpen
  \bibfield  {author} {\bibinfo {author} {\bibfnamefont {B.}~\bibnamefont
  {Schenke}}, \bibinfo {author} {\bibfnamefont {P.}~\bibnamefont {Tribedy}}, \
  and\ \bibinfo {author} {\bibfnamefont {R.}~\bibnamefont {Venugopalan}},\
  }\href {\doibase 10.1103/PhysRevLett.108.252301} {\bibfield  {journal}
  {\bibinfo  {journal} {Phys.Rev.Lett.}\ }\textbf {\bibinfo {volume} {108}},\
  \bibinfo {pages} {252301} (\bibinfo {year} {2012}{\natexlab{a}})},\ \Eprint
  {http://arxiv.org/abs/1202.6646} {arXiv:1202.6646 [nucl-th]} \BibitemShut
  {NoStop}%
\bibitem [{\citenamefont {Schenke}\ \emph
  {et~al.}(2012{\natexlab{b}})\citenamefont {Schenke}, \citenamefont
  {Tribedy},\ and\ \citenamefont {Venugopalan}}]{Schenke:2012hg}%
  \BibitemOpen
  \bibfield  {author} {\bibinfo {author} {\bibfnamefont {B.}~\bibnamefont
  {Schenke}}, \bibinfo {author} {\bibfnamefont {P.}~\bibnamefont {Tribedy}}, \
  and\ \bibinfo {author} {\bibfnamefont {R.}~\bibnamefont {Venugopalan}},\
  }\href {\doibase 10.1103/PhysRevC.86.034908} {\bibfield  {journal} {\bibinfo
  {journal} {Phys.Rev.}\ }\textbf {\bibinfo {volume} {C86}},\ \bibinfo {pages}
  {034908} (\bibinfo {year} {2012}{\natexlab{b}})},\ \Eprint
  {http://arxiv.org/abs/1206.6805} {arXiv:1206.6805 [hep-ph]} \BibitemShut
  {NoStop}%
\end{thebibliography}%

\end{document}